\documentclass[lettersize,twoside,journal]{IEEEtran}

\usepackage{graphicx}
\usepackage{amssymb}
\usepackage{amsmath,cases}
\usepackage{cite}
\usepackage{multirow}
\usepackage{color}
\usepackage[table]{xcolor}
\usepackage{booktabs}
\usepackage{balance}
\usepackage{acro}
\usepackage{enumitem}
\usepackage{caption}
\usepackage{subcaption}

\DeclareMathOperator{\Sspace}{\mathcal{S}}
\DeclareMathOperator{\Aspace}{\mathcal{A}}

\usepackage{color}

\definecolor{darkgreen}{RGB}{0,128,0}

\DeclareAcronym{AO}{
short = AO,
long = alternate optimization
}

\DeclareAcronym{BS}{
	short = BS,
	long = base station
}

\DeclareAcronym{BD-RIS}{
short = BD-RIS,
long = Beyond diagonal reconfigurable intelligent surface
}

\DeclareAcronym{CSI}{
	short = CSI,
	long = channel state information
}

\DeclareAcronym{D2D}{
	short = D2D,
	long = device-to-device
}

\DeclareAcronym{FP}{
short = FP,
long = fractional programming
}

\DeclareAcronym{IAB}{
	short = IAB,
	long = integrated access and backhaul
}

\DeclareAcronym{IoT}{
	short = IoT,
	long = Internet of Things
}

\DeclareAcronym{IoD}{
	short = IoD,
	long = Internet of Drones
}

\DeclareAcronym{ISCC}{
	short = ISCC,
	long = Integrated sensing\, communication\, and computation
}

\DeclareAcronym{KPI}{
	short = KPI,
	long = Key performance indicator
}

\DeclareAcronym{LoS}{
	short = LoS,
	long = line-of-sight 
}

\DeclareAcronym{LSTM}{
	short = LSTM,
	long = long short term memory
}

\DeclareAcronym{mMTC}{
	short = mMTC,
	long = massive machine type communication
}

\DeclareAcronym{ML}{
	short = ML,
	long = Machine learning
}

\DeclareAcronym{mmWave}{
	short = mmWave,
	long = millimeter wave
}

\DeclareAcronym{NTN}{
	short = NTN,
	long = non-terrestrial network
}

\DeclareAcronym{NW}{
	short = NW,
	long = Narrow band
}

\DeclareAcronym{nLoS}{
	short = NLoS,
	long = non line-of-sight
}

\DeclareAcronym{NLS}{
	short = NLS,
	long = non line-of-sight
}

\DeclareAcronym{PIN}{
	short = PIN,
	long = positive-intrinsic-negative
}

\DeclareAcronym{QNM}{
short = QNM,
long = quasi Newton method
}

\DeclareAcronym{RIS}{
short = RIS,
long = reconfigurable intelligent surface
}

\DeclareAcronym{RZF}{
short = RZF,
long = regularized zero forcing
}
\DeclareAcronym{SINR}{
	short = SINR,
	long = signal-to-interference-and-noise ratio
}

\DeclareAcronym{SNR}{
	short = SNR,
	long = signal-to-noise ratio
}

\DeclareAcronym{TN}{
	short = TN,
	long = terrestrial network
}

\DeclareAcronym{THz}{
	short = THz,
	long = Terahertz
}
\DeclareAcronym{URLLC}{
short = URLLC,
long = ultra-reliable and low-latency communication
}

\DeclareAcronym{UAV}{
	short = UAV,
	long = unmanned aerial vehicle
}

\DeclareAcronym{V2X}{
	short = V2X,
	long = vehicular-to-everything
}

\DeclareAcronym{WB}{
	short = WB,
	long = Wide band
}

\begin{document}
	
	\title{
		Multiconnectivity for SAGIN: Current Trends, Challenges, AI-driven Solutions, and Opportunities
	}
	\author{
		Abd~Ullah~Khan, Adnan~Shahid, Haejoon~Jung, and
		Hyundong~Shin
		\thanks{
			A. U. Khan, H. Jung, and H.~Shin (corresponding author)
			are with the Department of Electronics and Information Convergence Engineering, Kyung Hee University, Gyeonggi-do 17104 Korea; 
			A. Shahid is with ID Lab and imec, Ghent University, B-9000, Belgium.
		}
	}
	\markboth{
		Submitted for Publication in IEEE 
	}{ 
		Khan \textit{\MakeLowercase{et al.}}:
	}
	

	\maketitle

	\begin{abstract}
		Space-air-ground-integrated network (SAGIN)-enabled multiconnectivity (MC) is emerging as a key enabler for next-generation networks, enabling users to simultaneously utilize multiple links across multi-layer non-terrestrial networks (NTN) and multi-radio access technology (multi-RAT) terrestrial networks (TN). However, the heterogeneity of TN and NTN introduces complex architectural challenges that complicate MC implementation. Specifically, the diversity of link types, spanning air-to-air, air-to-space, space-to-space, space-to-ground, and ground-to-ground communications, renders optimal resource allocation highly complex. Recent advancements in agentic reinforcement learning (RL) have shown remarkable effectiveness in optimal decision-making in complex and dynamic environments. In this paper, we review the current developments in SAGIN-enabled MC and outline the key challenges associated with its implementation. We further highlight the transformative potential of learning-driven approaches for resource optimization in a heterogeneous SAGIN environment. To this end, we present a case study on resource allocation optimization enabled by agentic RL for SAGIN-enabled MC involving diverse RATs. Results show that agentic RL can effectively handle complex scenarios and substantially improve latency and capacity with reduced power consumption. Finally, open research problems and future directions are presented to realize efficient SAGIN-enabled MC.
		
	\end{abstract}
	
	
	
	\acresetall
	
	\section{Introduction}
	\label{sec:1}
	
	\IEEEPARstart {T}{he} 
	vision of sixth-generation (6G) wireless networks is driven by unprecedented key performance indicators (KPIs), including terabit-per-second data rates, end-to-end latencies below one millisecond, and `five-nines' availability, while supporting massive connectivity for billions of heterogeneous devices \cite{10918743}. Additionally, the integration of heterogeneous terrestrial networks (TNs) and non-terrestrial networks (NTNs) into a unified space–air–ground integrated network (SAGIN) for global coverage further intensifies these requirements \cite{10530195}. Meeting such ambitious requirements mandates fundamental innovations in network design, resource management, and connectivity paradigms.
	
	In this context, multiconnectivity (MC) has emerged as a promising solution that enables devices to maintain simultaneous links with multiple networks \cite{ZXJJHSG:25:IEEE_J_Net}. More precisely, MC enhances data rate through traffic aggregation across multiple networks. Similarly, it improves link reliability through traffic redundancy across various networks. Furthermore, MC reduces latency by using packet duplication and optimal path selection for data transmission. Likewise, MC supports load balancing by enabling dynamic bandwidth utilization and efficient resource coordination, redistributing traffic from congested nodes to underutilized ones. Accordingly, MC directly addresses the challenges of providing ubiquitous connectivity and 6G service classes, i.e., enhanced mobile broadband (eMBB), massive machine-type communication (mMTC), and hyper reliable low latency communication (HRLLC) across diverse environments \cite{M:24:IEEE_M_COM}.

	\subsection{MC in the Unified and Standardized SAGIN Architecture}
	
	The potential of MC is further amplified by the emerging unified and integrated architecture of SAGIN, 
	which spans multiple standardization bodies, including the 3rd Generation Partnership Project (3GPP) for 5G systems, the Consultative Committee for Space Data Systems (CCSDS) for space communications, and the Internet Engineering Task Force (IETF) for Internet Protocol (IP)-based networking \cite{YZMHMMAS:24:IEEE_J_JSAC}. Such an integrated approach enables seamless interoperability across heterogeneous platforms, allowing user devices and applications to dynamically access the most suitable link or network regardless of their location or mobility. By using standardized protocols and optimal coordination, such convergence leads to ubiquitous connectivity, efficient resource utilization, and resilient service provisioning, among others \cite{TJYLJHL:25:IEEE_J_CCN}.
	
\begin{figure*}[t!] 
		\centering
		\includegraphics[width=0.8\textwidth]{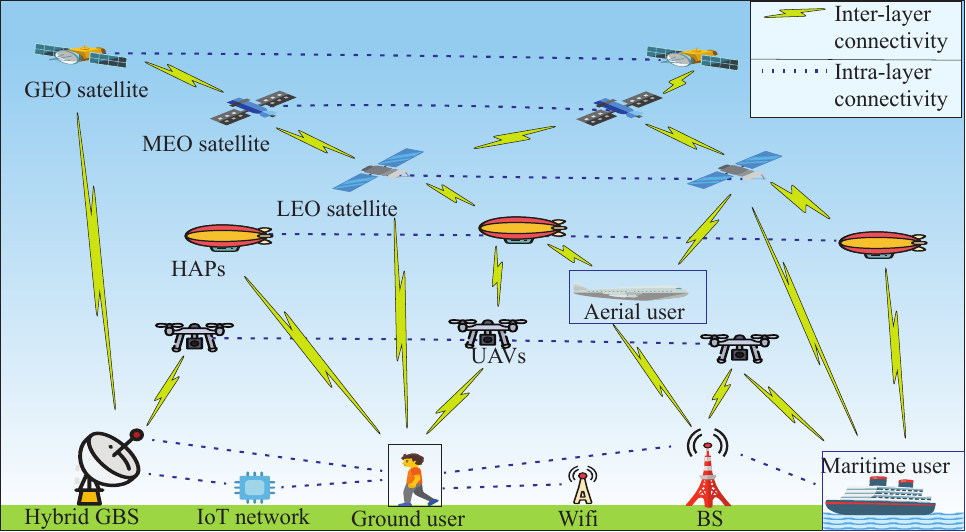}
		\caption{Overview of SAGIN-enabled MC.}
		\label{fig1}
	\end{figure*}

	Nonetheless, the integration of multi-layer NTNs and multi-RAT TNs introduces complex architectures and a plethora of use cases. Consequently, a variety of link types include air-to-air, air-to-space, space-to-space, space-to-ground, and ground-to-ground links. Fig. \ref{fig1} illustrates a scenario of the heterogeneous environment and architectures enabled by unified SAGIN, where the coexistence of space, air, and ground platforms is considered. As shown in the figure, it encompasses unmanned aerial vehicles (UAVs), high-altitude platforms (HAPs), satellites, and hybrid gateway-base station (GBS) with highly dynamic and complex network conditions, which makes it more challenging to optimize resource allocation and management. For instance, rapidly changing channel conditions, frequent handovers across space, air, and ground domains, and varying latency, bandwidth, and reliability requirements are the factors that contribute to the resource management complexity.
	
	Consequently, leveraging MC is more challenging in highly complex and heterogeneous network topologies and link types \cite{10909704}. Recently, reinforcement learning (RL) and agentic artificial intelligence (AI) have shown great potential for coping with highly dynamic and complex environments as observed in the unified SAGIN \cite{10500738, JKTHLX:24:IEEE_J_MCOM}. In particular, RL-driven agentic AI, called agentic RL, can be leveraged to intelligently and efficiently perform link selection for MC in such environments. In this paper, we explore MC in the emerging SAGIN paradigm, highlighting the connected challenges and the AI potential in realizing a unified AI-driven SAGIN-enabled MC. The novelty and contributions are summarized as follows. 
	
	\begin{itemize}
		\item We explore MC in the emerging unified SAGIN paradigm, outlining its current development and standardization status. Additionally, we investigate the critical challenges associated with designing SAGIN-enabled MC systems. 
		
		\item We highlight the potential of agentic RL in addressing the complex challenges posed by SAGIN-enabled MC and propose an agentic RL-based algorithm for optimizing MC resources that integrates multi-layer NTNs with multi-RAT TNs. The proposed agent continuously senses link states, plans link combinations, and acts over time with memory and explicit goals.
		This algorithm employs a multi-objective agentic framework and optimizes link selection by jointly considering three critical KPIs, such as capacity, latency, and power consumption. To safeguard against instability in the agent's learning process caused by the highly dynamic SAGIN environment and to ensure policy convergence, we incorporate a stabilization mechanism.

		\item The proposed algorithm is validated through simulations and compared with both model- and learning-driven baselines. In addition, we present a comprehensive overview of open challenges and potential research directions. 	
		
	\end{itemize}

	\section{Standardization and Challenges}
	\subsection{Standardization}
	The 3GPP introduced the MC concept in Releases 11 and 12 to enhance link diversity and robustness \cite{YZMHMMAS:24:IEEE_J_JSAC}. Releases 15 and 16 further improved the architecture and protocols. Release 17 integrated NTN platforms, such as low Earth orbit (LEO), medium Earth orbit (MEO), and geostationary Earth orbit (GEO) satellites, and HAPS, into 5G. It also standardized New Radio (NR)-based satellite access in the S- and L-bands, which facilitated transparent payload configurations and frequency division duplexing for handheld devices. 
	Release 18 included support for Ka-band to access NTN in MC. Release 19 is expected to support MC across 5G NR, NTN, and non-public networks, supporting seamless handover across heterogeneous cells.
	In addition, Release 19 supports AI techniques for intelligent and context-aware resource management and allocation in MC \cite{X:25:IEEE_M_CSM}.
	\subsection{Challenges}
	Although SAGIN-enabled MC has the potential to provide ubiquitous, resilient, and high-capacity connectivity, there are several challenges to its real-world implementation, some of which are described below.

	\subsubsection{Dynamic 3D Network Topology}
	The spatial distribution of SAGIN nodes, including LEO/MEO/GEO satellites, HAPs, UAVs, and terrestrial BSs, forms a dynamic three-dimensional (3D) topology. Traditional routing and handover models are inefficient as they are designed for static, planar, and homogeneous network architectures. Achieving continuously adaptive path planning, timely channel state information (CSI) acquisition, and real-time resource management is challenging in SAGIN due to its highly dynamic 3D topology, frequent link variations, and heterogeneous network layers. These factors make real-time coordination and information acquisition extremely complex, resulting in frequent link disruptions.

	\subsubsection{Signal Attenuation, Scattering, and Doppler Shift}
	TN signals experience attenuation, multipath propagation, and scattering in dense urban areas. Similarly, satellite signals, particularly those in the Ka and V bands, undergo rain fade. Moreover, the mobility of LEO satellites introduces Doppler shifts and necessitates frequent handovers by users. These impairments, induced by propagation and mobility, lead to frequent link drops and handovers, increased signaling overhead, packet loss, and latency. They also create synchronization problems, resulting in inefficient resource utilization and management. Consequently, maintaining stable, synchronized, and high-capacity multiconnectivity becomes challenging.

	\begin{figure*}[h] 
		\centering
		\includegraphics[width=0.8\textwidth]{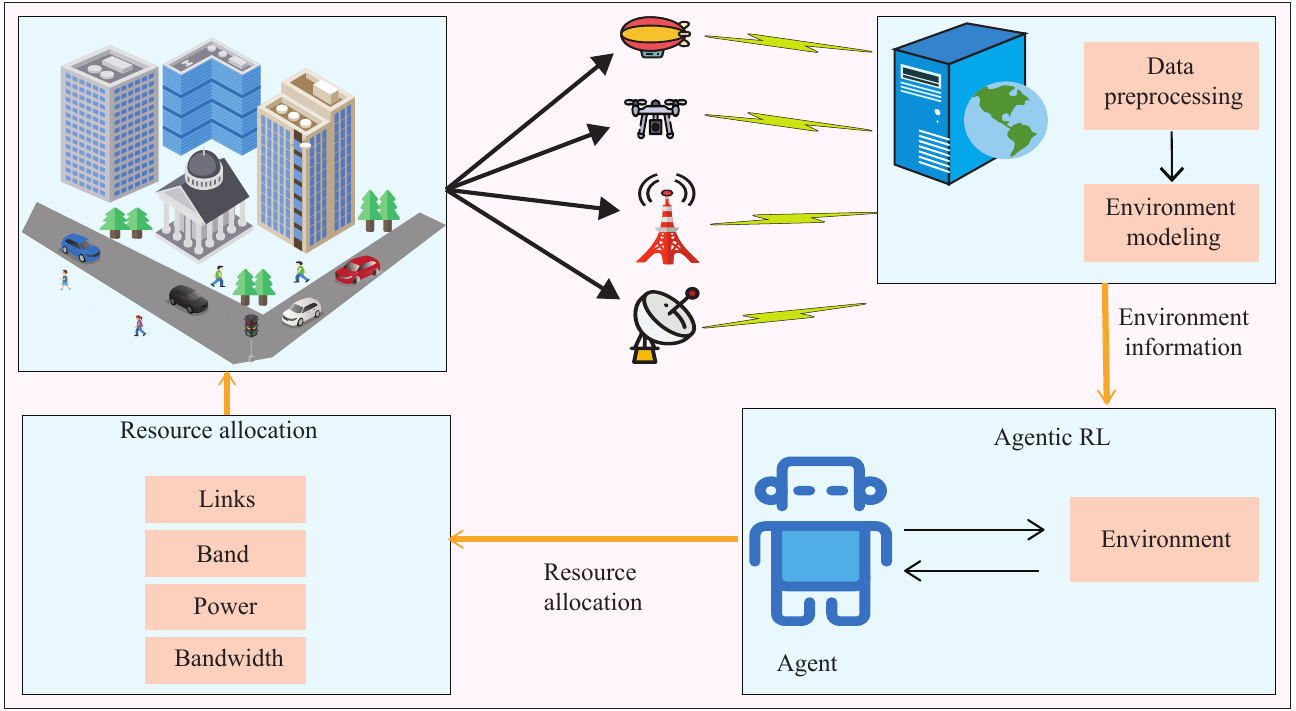}
		\caption{The high-level depiction of the proposed agentic RL framework for resource allocation in SAGIN-enabled MC.}
		\label{fig-system-architect}
	\end{figure*}

	\subsubsection{Efficient Scheduling and Protocol Stack Adaptation}
	The scheduler for SAGIN-enabled MC needs to differentiate between diverse quality-of-service (QoS) classes, .e.g, HRLLC, eMBB, etc. However, the drastic differences between TN and NTN nodes in jitter, delay, packet loss, and cost, make dynamic QoS mapping to meet the service requirements challenging. Similarly, the scheduler must dynamically reconfigure the protocol stack layers (e.g., splitting the packet data convergence protocol (PDCP) for multi-path transmission) and adapt them in real time to meet the various QoS requirements. Efficiently scheduling heterogeneous traffic (i.e., HRLLC, eMBB, and mMTC) in a highly dynamic SAGIN environment is challenging for an SAGIN-enabled MC.


	\section{Agentic RL for SAGIN-enabled MC}
	Recently, agentic RL has emerged as a key tool for addressing complex network optimization challenges \cite{jiang2025verltool}. In the context of SAGIN-enabled MC, such agents optimize resource allocation under dynamic and complex conditions. The agents continuously learn rapidly changing network states, diverse user requirements, and complex mobility patterns. This enables them to adapt to environmental dynamics and facilitate efficient service delivery. The agents also enable proactive decision-making regarding link activation, handover, traffic steering, and load balancing to predict network conditions. This allows for the intelligent definition of dynamic resource allocation patterns and the optimization of parameters such as transmit power, bandwidth, and carrier selection to maximize spectral and energy efficiency. Thus, agentic RL can enable intelligent, self-optimizing control policies for SAGIN-enabled MC. These policies can orchestrate handovers, link selection, resource allocation, and load balancing across heterogeneous domains while ensuring scalability and adaptability under highly nonlinear dynamics.


	\begin{figure*}[htbp] 
		\centering
		\includegraphics[width=0.8\textwidth]{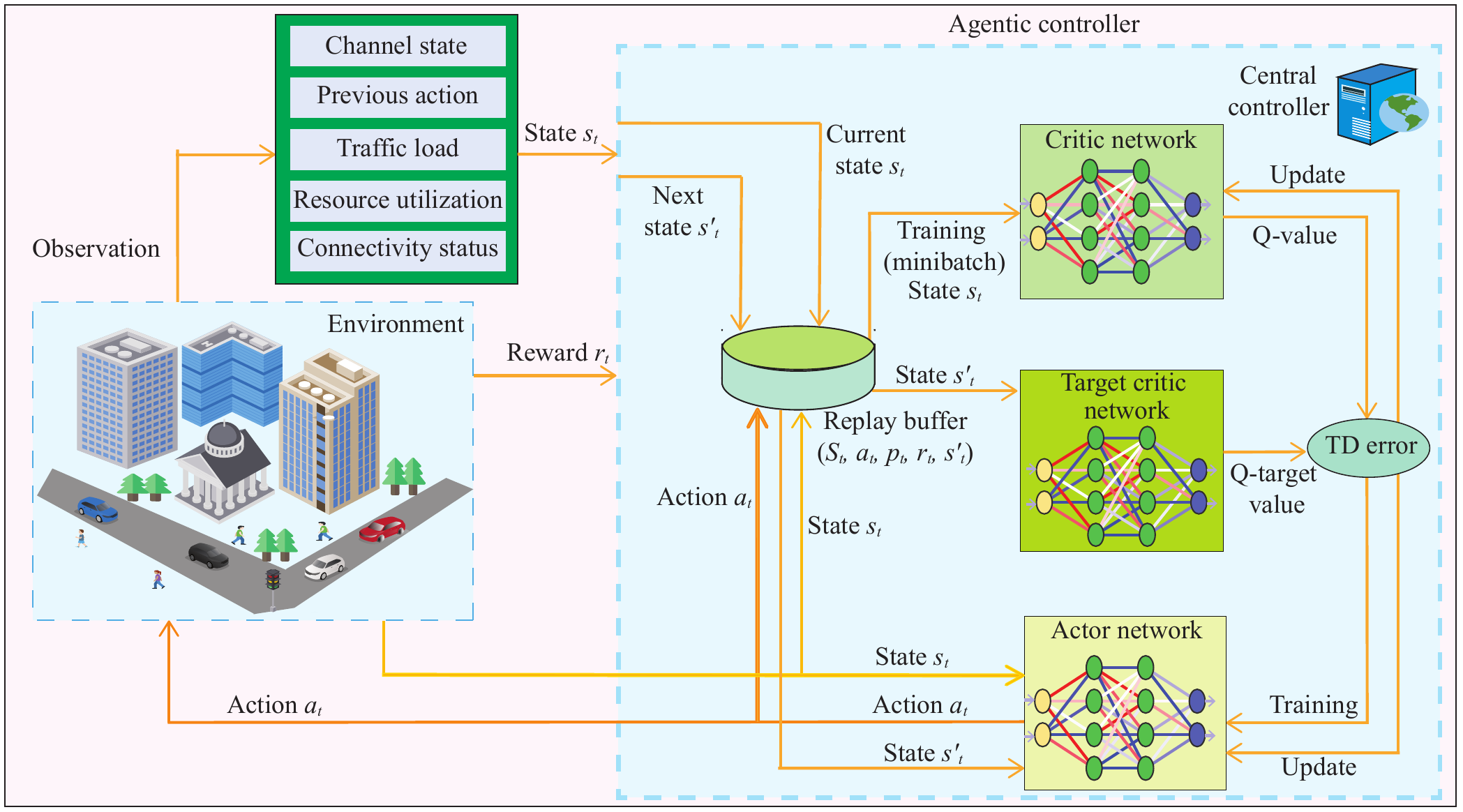}
		\caption{System architecture of the proposed algorithm. The agent performs perception (environment modeling, data preprocessing), planning/decision (actor–critic), and memorization (replay buffer, target critic).}
		\label{fig-actor-critic}
	\end{figure*}
	
	
	To demonstrate its potential for SAGIN-enabled MC, we propose an agentic RL-based algorithm that optimizes link selection and scheduling within the complex and dynamic environment of SAGIN-enabled MC. The objective is to select the optimal set of links for each user equipment (UE) to satisfy its quality of service (QoS) requirements.  To this end, we formulate QoS fulfillment on multiple links as a partially observable Markov decision process. In this process, the agent can only observe the state information of the link(s) it has selected at any given observation time. 
	
	Consequently, we have $\{\Sspace,\Aspace,p(s'_t |  s_t,a_t ),r_t\}$, where $\Sspace$ denotes the state space of the environment, while $\Aspace$ represents the action space for making the optimal decision. In addition, $p(s'_t | s_t,a_t )$ corresponds to the probability of transitioning to state $s'_t\in \Sspace$ conditioned on taking action $a_t\in \Aspace$ in state $s_t\in \Sspace$. The term $r_t$ represents the immediate reward obtained by taking action $a_t$, which drives the transition from $s_t$ to $s'_t$. We assume that a UE has access to $N$ available links, indexed by $n \in\{1,..., N\}$, from which a subset can be dynamically assigned to the UE.

	\subsection{State Space} The state of each selected channel combination is modeled as a binary state variable that indicates its ability to support the required minimum QoS in terms of capacity, latency, and power consumption. Specifically, for a combination of $N$ links at time $t$, a state $s_{nt}$ is assigned a value of $+1$ if the combination is available (AV) and capable of meeting the minimum QoS requirements. On the other hand, it is assigned a value of $-1$, if it is not available (NAV).
		Due to the dynamic environment, the state of each of the N links changes dynamically, i.e., the availability of QoS between AV and NAV changes. We assume the maximum number of available link combinations at any given time is $N = 4$. The link state space is then defined as the set of binary state vectors as $S = \{s_t  | s_t  = [s_{1t},...,s_{Nt}]\}$, which subsequently yields a state space of the form {\small $S = \{[AV, AV, AV, AV], ..., [NAV, NAV, NAV, NAV]\}$}. It is noted that the AV and NAV states of a platform are determined by three factors: CSI from the platform, the platform's load, and the traffic type of each UE that reflects its QoS requirements.

		\subsection{Actions} An action is defined as the selection of a subset of the $N$ combinations. The action space is defined as the set of binary vectors $\Aspace = \{a_t | a_t = [C_{1t}, ..., C_{Nt}]\}$, where $C_{nt} = 1$ shows that the $n$-th link combination is selected at time $t$, while $C_{nt} = 0$ means the opposite (i.e., not selected).

		\subsection{Reward} The immediate reward $r_t$ is formulated as a penalty imposed on the agent that is proportional to the capacity, latency, and power usage calculated over an episode. More precisely, the reward at each time step is defined as a weighted combination that increases with higher capacity and decreases with higher latency and power consumption. Thus, the reward function maximizes the achievable transmission rate while minimizing system delay. The function implicitly penalizes packet loss, high delays, and exceeding network load constraints, thereby guiding the agent to learn optimal link selection policies. Furthermore, it encourages selecting fewer links to minimize cost.

		\subsection{Agent Architecture}
		As illustrated in Fig. \ref{fig-actor-critic}, the proposed learning agent is tailored for	SAGIN-enabled MC decision-making, where link availability, traffic dynamics, and resource utilization jointly evolve across heterogeneous space–air–ground layers. 
        The agent architecture comprises three neural networks: the actor $A(s_t)$, the critic $C(s_t, a_t)$, and the target critic $TC(s_t, a_t)$.
		
		\begin{itemize}
			\item The actor learns a stochastic policy $\pi$ that maps the agent's observation of the state i.e., channel state, traffic load, resource utilization, connectivity status, and previous action to the action space $a_t$ representing link selection and scheduling decisions across available SAGIN links.
			Specifically, the actor determines which subset of satellite, HAP, UAV, and terrestrial links should be activated at time $t$ to satisfy QoS requirements. The policy is optimized to maximize the expected long-term cumulative reward,
				$\pi^*(a_t|s_t) = \arg\max_{a_t} \mathbb{E} \left[ \sum_{t=0}^{\infty} \gamma^t r_t \right]$,
			where $\gamma$ is the discount factor that controls the contribution of future rewards. Also, $\mathbb{E}[\cdot]$ denotes the expected value operator.
			
			\item The critic estimates the state-action value, quantifying the quality of the actor's
				link-selection and scheduling decisions in terms of jointly optimized KPIs, i.e., capacity,
				latency, and power consumption. This way, the critic guides the actor towards policies that are robust to frequent topology changes and traffic fluctuations in SAGIN.
				\item To stabilize learning in the highly dynamic SAGIN environment, a target-critic is introduced to compute the Bellman equation-based estimates of future state-action values. 
				After action $a_t$ is taken, i.e., utilization of the selected MC links started, the environment returns the immediate reward $r_t$ and the next state $s'_t$ reflecting updated link conditions and resource availability. The target-critic estimates future state-action values as the sum of the immediate reward and the discounted value of the next state, thereby mitigating oscillations caused by rapidly varying	channel and mobility conditions.
			
		\end{itemize}

		\subsection{Network Update}
		The temporal-difference (TD) error is calculated as the difference between the obtained reward plus the discounted estimated future value and the current estimated value of the state-action pair \cite{10500738}. The critic network is updated by minimizing the squared TD error to ensure that its estimated Q-values closely match the target values. The actor network is updated using the policy gradient method, wherein the gradient is computed based on the TD error to adjust the policy to maximize the expected cumulative reward.

		\section {Case Study of the SAGIN-enabled MC}
		In this section, we present a case study demonstrating the effectiveness of the proposed approach for optimizing SAGIN-enabled MC. To this end, we conduct a simulation in a stochastic SAGIN environment comprising ground, aerial, and space platforms. 
        The agent and scheduler are deployed at a centralized controller in GBS, where link-selection decisions are made based on channel availability and quality, previous actions, platform load, and resource utilization. Each episode consists of 50 time-steps, during which an agent selects any non-empty subset of the four candidate links from BS, UAV, HAP, and LEO satellite. We model the line-of-sight (LOS) probability of a link using standard propagation formulations from 3GPP TR 38.901 \cite{7503971} and ITU-R P.1410/P.618 \cite{10999379}, as well as geometry-based models, as in \cite{6863654} and \cite{7414036}. 
		
		For performance evaluation, we consider capacity, latency (i.e., the sum of propagation delay and service time), and power consumption through per-link power consumption. Unless otherwise stated, the reward weights of $(1.0, 0.2, 0.05)$ are used to jointly optimize capacity, latency, and power consumption, respectively. These weights are selected through empirical tuning to achieve a reasonable tradeoff among the KPIs and ensure stable learning. The other simulation parameters are listed in Table \ref{tab:sim_parameters}. 
		
		\begin{table}[t!]
			\small
			\centering
			\caption{Simulation Parameters}
			\renewcommand{\arraystretch}{1.05}
			\setlength{\tabcolsep}{3.2pt}
				\begin{tabular}{l l}
					\toprule
					\textbf{Parameter}  & \textbf{Value / Description} \\
					\midrule
					Max. number of links & 4 (BS, UAV, HAP, LEO) \\
					Bandwidths (MHz)  & BS 100; UAV 200; HAP 200; LEO 250 \\
					Frequency (GHz)  & BS 28; UAV 26; HAP 26; LEO 27 \\
					Transmit power (dBm)  & BS 30; UAV 27; HAP 35; LEO 40 \\
					Power cost (W)  & BS 2; UAV 3; HAP 4; LEO 5 \\
					UE height & 1.5 m \\
					BS location / height  & $(200,150,25)$ m \\
					UAV speed  & 15 m/s (random waypoint) \\
					UAV altitude range  & 120--250 m \\
					HAP altitude  & 20 km \\
					LEO altitude  & 550 km (circular arc motion, 4$^{\circ}$/step) \\
					Urban cell radius  & 500 m \\
					Learning rate  & $1\times10^{-3}$ \\
					Discount factor  & 0.99 \\
					Entropy regularization   & 0.01 \\
					Episodes $\times$ steps  & 10000 episodes $\times$ 50 steps \\
					\bottomrule
				\end{tabular}
				\label{tab:sim_parameters}
			\end{table}
			
			\begin{figure*}[htbp]
				\centering
				\begin{subfigure}[b]{0.40\textwidth}
					\includegraphics[width=\textwidth]{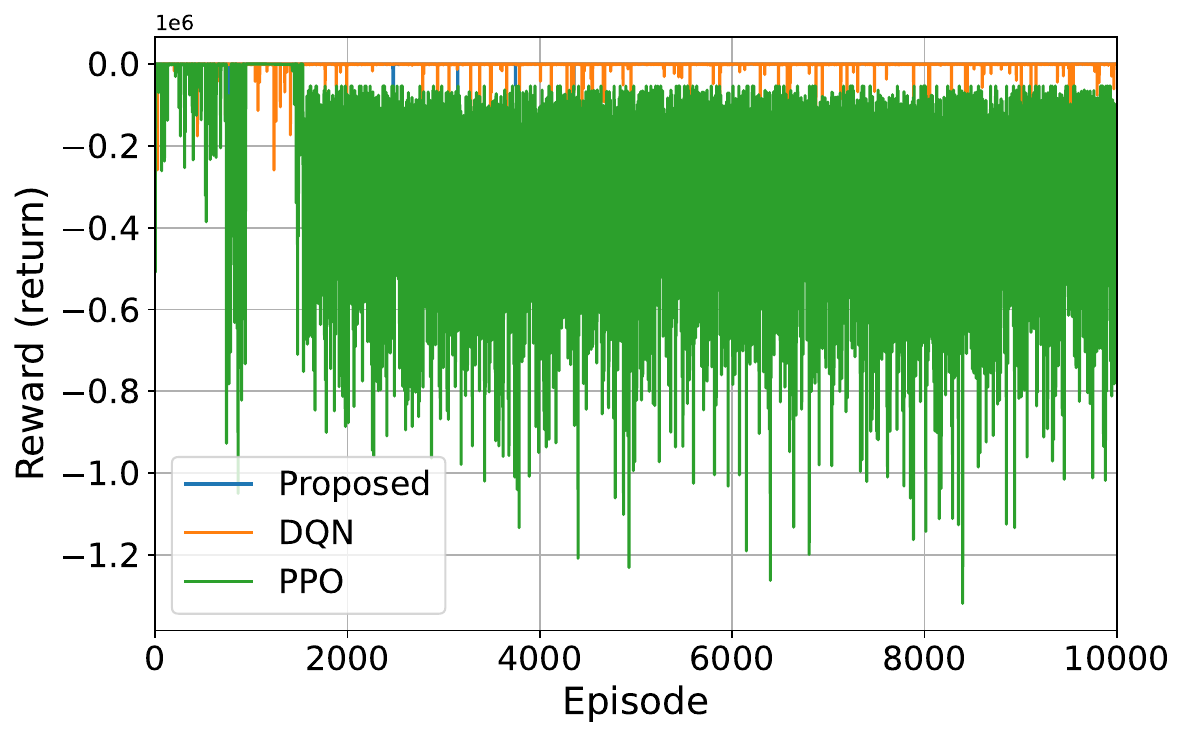}
					\caption{}
					\label{fig:Return}
				\end{subfigure}
				\begin{subfigure}[b]{0.40\textwidth}
					\includegraphics[width=\textwidth]{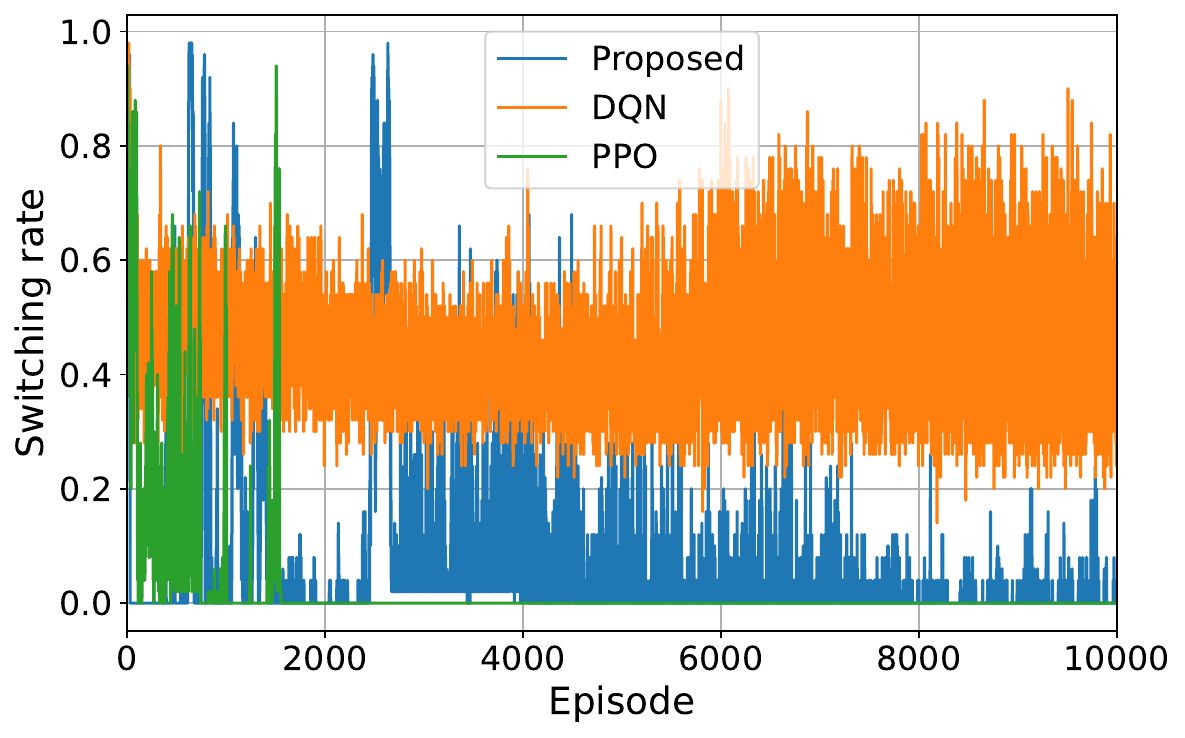}
					\caption{}
					\label{fig:Switch}
				\end{subfigure}
				\begin{subfigure}[b]{0.40\textwidth}
					\includegraphics[width=\textwidth]{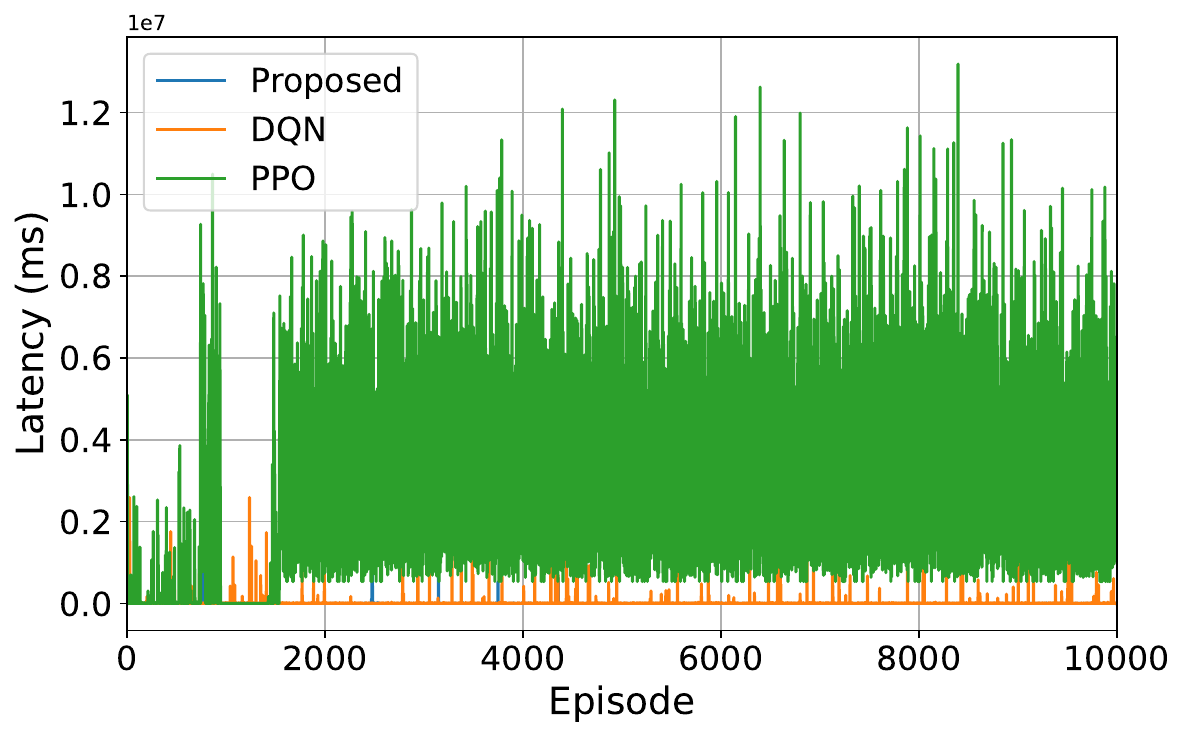}
					\caption{}
					\label{fig:Latency}
				\end{subfigure}
				\begin{subfigure}[b]{0.40\textwidth}
					\includegraphics[width=\textwidth]{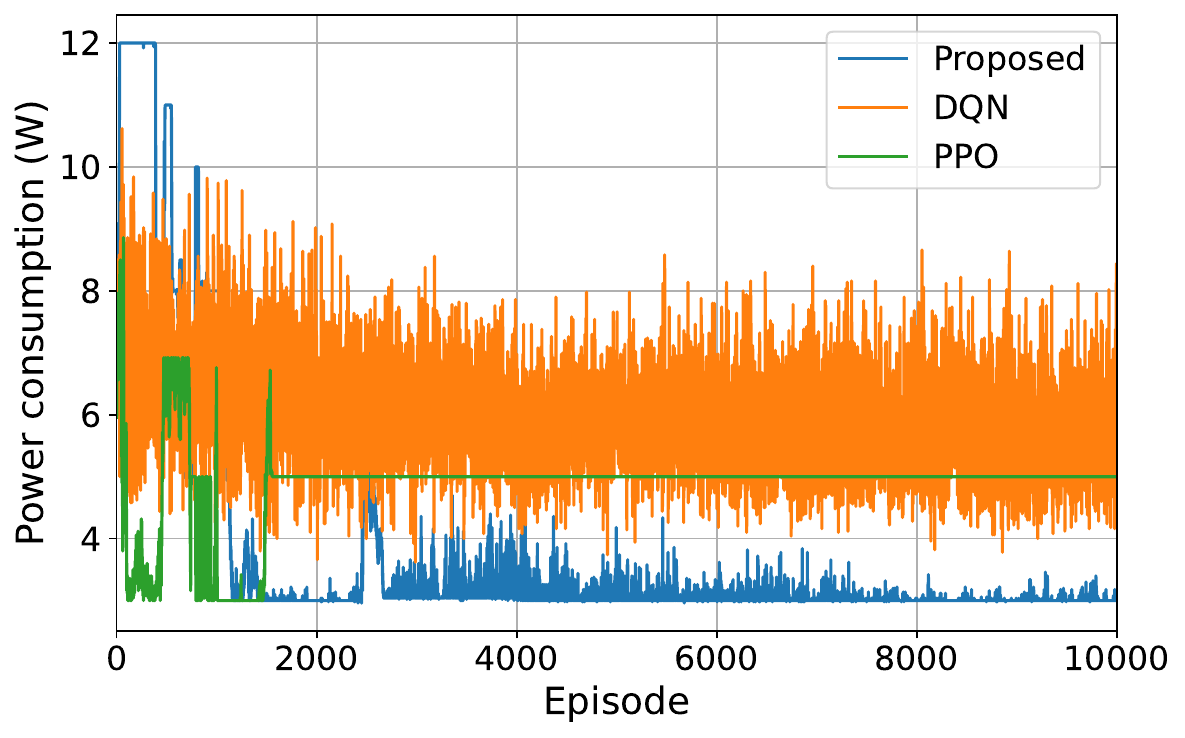}
					\caption{}
					\label{fig:Power}
				\end{subfigure}
				\caption{The illustration of the learning behavior and convergence performance of the agent: (a) episode reward, (b) switching rate, (c) average latency, and (d) power consumption.}
				\label{fig:main}
			\end{figure*}

			\subsection{Learning Behvaior and Convergence}
			Fig. \ref{fig:Return} shows the total reward/return plotted over training episodes for the SAGIN-enabled MC. The proposed algorithm converges rapidly compared to DQN and PPO, which is attributable to its stabilizing components that keeps the policy updated in highly dynamic environments.
			DQN exhibits slower convergence and higher variability due to its off-policy nature, which cannot adapt efficiently to the highly dynamic SAGIN environment.
			Similarly, the inefficient optimization policy of PPO in the highly dynamic SAGIN environment limits its ability to fully exploit favorable link combinations, resulting in a lower return.

			Fig. \ref{fig:Switch} illustrates the links switching rate among the available links during each episode.
		The proposed algorithm increasingly reduces the switching rate, converging toward near-zero values, as training progress. DQN consistently maintains high switching rate, indicating it struggle with optimal links selection in the highly dynamic SAGIN environment. PPO shows the lowest switching rate. However, given its low return values, this lowest switching rate can be interpreted as locking to fixed and sub-optimal links selection, which results in low return.
			
			In Fig. \ref{fig:Latency},  		the proposed scheme consistently achieves lower end-to-end latency. DQN performs better than PPO. The conservative updates in PPO prevent it from exploiting low-latency terrestrial and aerial links when conditions are favorable.
			Similarly, Fig. \ref{fig:Power} shows the average transmit power in watts. 
			The proposed method maintains lower power usage than both DQN and PPO. DQN utilizes high-power aerial and satellite links during exploration, leading to persistently higher energy expenditure. PPO performs better but still incurs  higher power usage due to its slower policy adaptation.

			\subsection{Performance Evaluation}
			We compare the proposed algorithm against random link selection, round-robin single-link cycling, greedy single-link selection based on the highest signal-to-noise ratio (SNR), BS-only conservative terrestrial baseline, DQN, and PPO.
			
			\subsubsection{Capacity}
			Fig. \ref{fig-Capacity_comp} depicts the average capacity performance. The proposed scheme excels all the competing schemes due to intelligently selecting the optimal set of ground, aerial, and space links simultaneously. The single-best (based on the highest SNR) scheme stands as the second-best performing, however, it is limited by single-link selection. The random and round-robin approaches suffer from inefficient link usage and suboptimal selections, resulting in significantly lower capacities. The single-BS configuration yields the lowest capacity as it lacks access to the additional spectral and spatial resources of UAV, HAP, and LEO platforms. DQN performs better than random and round-robin; however, its performance is limited by its limited action representation and suboptimal exploitation of multiple links. PPO performs lower due to its conservative policy updates and higher sensitivity to delayed rewards in highly heterogeneous SAGIN environments.
			
			\subsubsection{Latency}
			In Fig. \ref{fig-Latency_comp} compares the average end-to-end latency. The proposed scheme achieves the lowest latency by intelligently selecting low-delay ground and aerial links. 
			DQN achieves low latency compared with other schemes. 
			In contrast, PPO experiences high latency.
			The Random and Round-Robin schemes incur delays that are several orders of magnitude higher due to inefficient link utilization and the selection of long-distance or weak channels. 
			The single-best scheme performs moderately well, yet it is suboptimal as high-SNR links, particularly NTN links, still incur large propagation delays. 
			The single-BS scheme yields stable yet higher latency than the proposed scheme as it lacks multi-link connectivity.

			\subsubsection{Power Consumption}
\begin{figure*}[t]  
				\centering
				
				\begin{subfigure}[b]{0.32\textwidth}
					\centering
					\includegraphics[width=\linewidth]{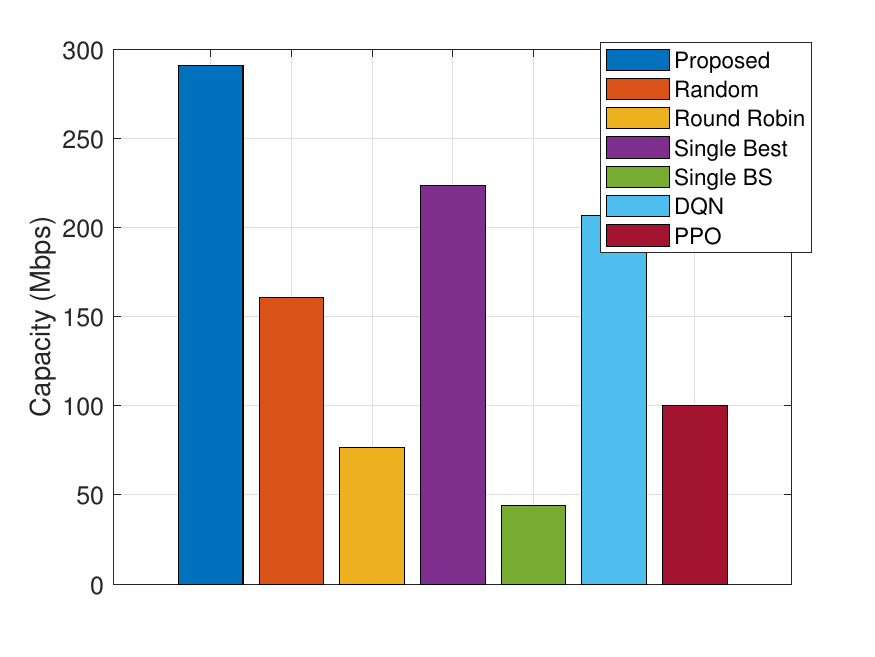}
					\caption{}
					\label{fig-Capacity_comp}
				\end{subfigure}
				\hfill
				\begin{subfigure}[b]{0.32\textwidth}
					\centering
					\includegraphics[width=\linewidth]{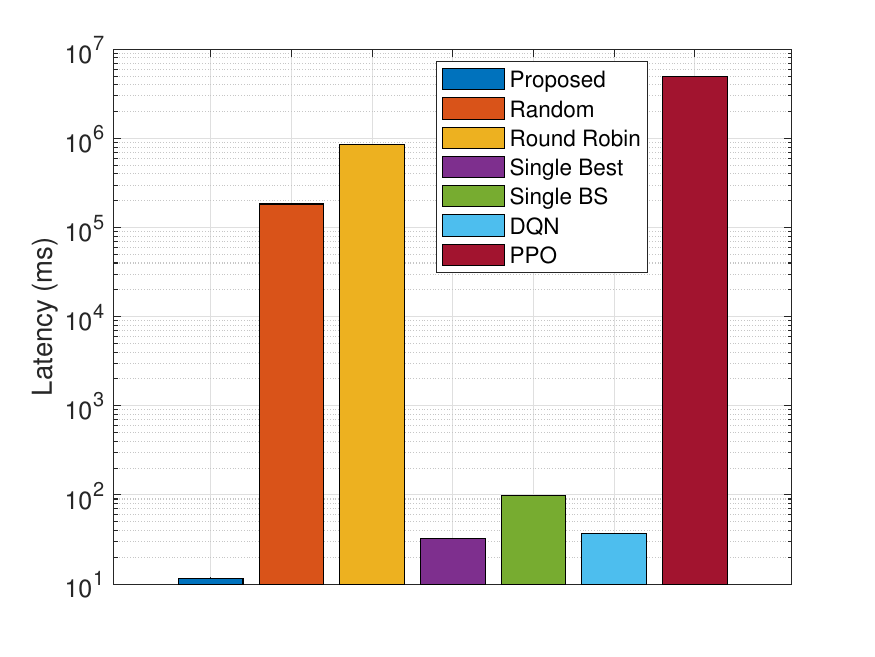}
					\caption{}
					\label{fig-Latency_comp}
				\end{subfigure}
				\hfill
				\begin{subfigure}[b]{0.32\textwidth}
					\centering
					\includegraphics[width=\linewidth]{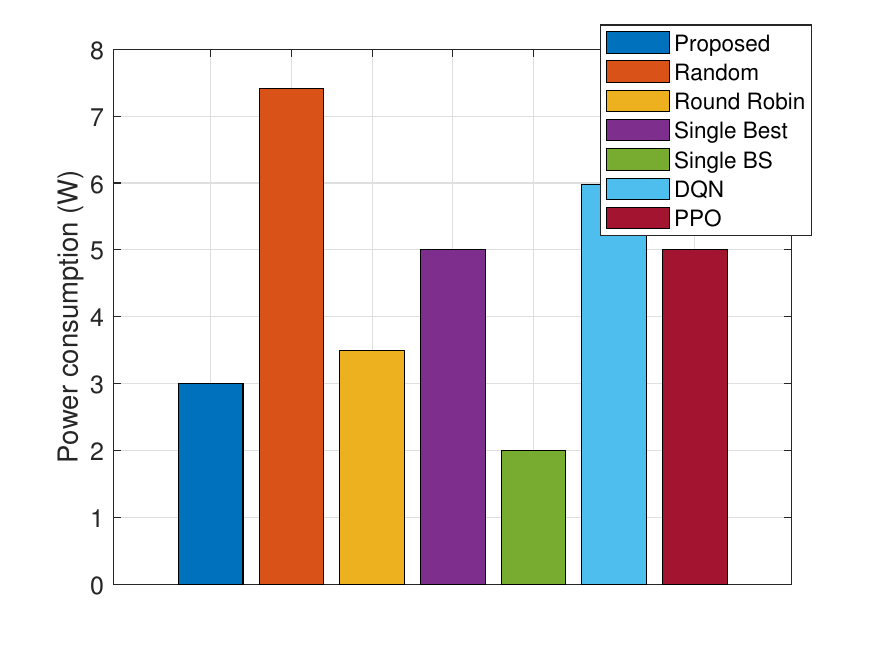}
					\caption{}
					\label{fig-Power_comp}
				\end{subfigure}
				
				\caption{Performance comparison of the proposed scheme with baseline methods in terms of (a) capacity, (b) latency, and (c) power consumption.}
				\label{fig:all_three}
			\end{figure*}

			Fig. \ref{fig-Power_comp} compares average power consumption. 
			The proposed scheme has a higher average power cost than single-BS because of using multiple links. The single-best and single-BS schemes also consume comparatively low power as they rely on only one active link. The random and round-robin methods consume an intermediate amount of power. 
			DQN shows moderate power usage, while PPO consumes comparatively higher power due to its less stable policy behavior and inefficient link activation.

			\section{Open Problems and Future Research Directions}
			This section discusses open issues and potential future research directions to improve the design of the SAGIN-enabled MC.

			\subsection{Synchronization}
			Perfect end-to-end coordination in SAGIN among sub-systems that differ in radio, mobility, transport, core, and backhaul, is crucial for optimal resource utilization. However, the variable and asymmetric propagation delays especially in satellite links disrupt time and frequency synchronization	required for coordinated transmissions. 
			For example, the round-trip delay from a GEO to a UE (about $500$ ms) is considerably higher than from a ground BS to a UE (typically less than $10$ ms) or from a LEO to a UE (about $25$ to $50$ ms). This discrepancy disrupts the congestion control and retransmission policies of well-established protocols such as TCP. 
			
			This problem can be solved by creating adaptive versions of existing transport protocols that decouple satellite and ground platforms. This ensures that long-delay satellite paths do not affect the responsiveness of ground-based flows. Furthermore, time-stamping and clock recovery techniques, along with delay compensation, can be developed to coordinate scheduling, transmission timing, and packet ordering accuracy. Similarly, dynamic buffer management can balance capacity and latency across paths with varying round-trip times.
			
			Application placement at ground, aerial, or space nodes can mitigate latency and jitter and achieve near real-time synchronization. Aside from this, standardized interfaces and unified control protocols, such as open and interoperable control through radio access network (RAN) intelligent controllers in open-RAN can be explored to achieve synchronization across heterogeneous domains in SAGIN. Additionally, synchronization is hindered by uplink power limitations for strong satellite connections on UEs, creating throughput asymmetries. These can be overcome by considering service-aware uplink scheduling and smart relay nodes (e.g., UAVs) for intermediate relaying.

			\subsection{Integrated Sensing and Communication (ISAC)}
			ISAC opens up a plethora of opportunities for SAGIN-enabled MC by enabling the joint use of communication and sensing capabilities across space, air, and ground platforms. As a result, in addition to transmitting data, nodes perform environmental awareness, making the network more context-aware in terms of resource allocation and interference avoidance. Furthermore, ISAC enables precise mobility tracking by extracting range, angle, Doppler, and trajectory information from communication waveforms. It also supports cooperative perception, which allows multiple heterogeneous nodes (e.g., satellites, UAVs, and BSs) to fuse their sensing observations and overcome individual sensing blind spots. Additionally, ISAC creates a data-rich environment that enhances AI-driven orchestration and predictive decision-making across SAGIN.
			
			However, ISAC introduces tradeoffs between the bandwidth used for sensing and communication. This trade-off can be mitigated by developing a coherent joint waveform design and sensing-aware scheduling across various SAGIN platforms, which is challenging to achieve. A promising research direction in this context is to use RL algorithms supplemented by generative AI models to proactively predict the necessary waveform and sensing parameters to efficiently operate in the complex, dynamic SAGIN environment.

			\subsection{Security and Privacy}
			SAGIN-enabled MCs substantially expand the threat domains of security and privacy. With its multiple interfaces and heterogeneous links spanning TN and NTN platforms, SAGIN-enabled MC introduces more vulnerabilities. Existing encryption and authentication schemes are designed primarily for homogeneous terrestrial networks and do not account for this type of multifaceted security and privacy risk. Link-aware, end-to-end encryption schemes can be designed for SAGIN-enabled MC such that authentication works seamlessly across TN and NTN platforms. In this context, quantum-resistant encryption is a promising long-term solution for protecting SAGIN systems against emerging threats. Similarly, sensing-based authentication is a new security paradigm enabled by ISAC, whereby physical-layer sensing features such as channel characteristics, mobility signatures, and environmental reflections can be used to verify identities and detect spoofing attacks. This paradigm can be considered for SAGIN-enabled MC.


			\subsection{Hardware and Software Complexity}
			In order to support SAGIN-enabled MC, UE must be able to operate at different frequency bands, power levels, and modulations. Similarly, they must adapt to various protocols featuring variations in framing, addressing, and control signaling. Efficient baseband co-design is required to allow for the shared processing of modulation, coding, synchronization, and resource allocation tasks. Additionally, software-defined radio can be used to dynamically select network parameters, such as waveforms, frequency bands, and medium access protocols, via software updates.
			
			Aside from this, modular hardware design should be considered to ensure hardware scalability, cost-effectiveness, interoperability, and maintainability. This will enable UEs to adaptively perform in dense urban areas where both TN and NTN links exist, as well as in remote rural areas where only NTN links exist. Furthermore, SAGIN-enabled MC operations are energy-intensive and require optimization techniques. Sleep-mode optimization can be developed across various network entities to enhance energy efficiency. Additionally, beam-triggered wake-up can be used to activate beams and subsystems on demand and deactivate idle beams and components. Additionally, green scheduling, in which energy efficiency is a primary objective of resource allocation, can be explored.

			\subsection{Resource Allocation and Sharing}
			SAGIN-enabled MC can benefit greatly from dynamic spectrum sharing and power allocation among space, air, and ground platforms. However, imperfect synchronization and coordination lead to conflicts in resource utilization under such a shared spectrum paradigm, creating co-channel interference and fairness concerns. Additionally, environmental variations, such as asymmetric rain fade, introduce nonlinear signal attenuation in high-frequency aerial and space links. This leads to fluctuating signal-to-interference-plus-noise ratio (SINR), frequent resource reassignments, and power imbalances across SAGIN platforms.
			
			One promising approach to addressing the synchronization challenge is designing a hierarchical coordination framework. In this framework, dedicated controllers for the satellite, aerial, and terrestrial layers operate under the supervision of a global controller, which aligns their timing and resource policies. Similarly, designing AI-driven predictive coordination models that anticipate link or load variations and proactively adjust spectrum and power allocation is an interesting research problem to explore.
			
			To offset losses caused by environmental dynamics, link control mechanisms such as adaptive modulation, coding rate adaptation, dynamic beamforming, and power control can compensate for the resulting losses. Additionally, multi-path diversity techniques can be developed to reroute or duplicate data through aerial or ground links when satellite links degrade due to rain. These techniques enhance robustness. Similarly, AI-based, weather-aware resource allocation and scheduling techniques can proactively reassign resources before a fade occurs, thereby maintaining service continuity.
			AI-based signal processing techniques combined with predictive mobility models can also mitigate Doppler-induced impairments and handover-related signaling overhead. Furthermore, signaling overhead can be reduced by AI-based node-level autonomy, such as edge computing and distributed AI, which enables local information processing and real-time decision-making.

			\subsection{Billing and Monetization Models}
			Billing and monetization in SAGIN-enabled MC are more complex because the traditional single-operator model is not applicable. Ideally, users would be billed based on their real-time usage across several network domains. However, accurately measuring and attributing data consumption for each network segment and operator becomes challenging in  SAGIN-enabled MC. One way to address this issue is to implement a centralized accounting system that continuously monitors network resource utilization and QoS provisioning indicators. These systems require cross-operator coordination through standardized application programming interfaces (APIs) that allow service providers to reconcile usage records, update pricing based on resource consumption, and provide users with transparent, verifiable billing statements. Another technical solution for ensuring billing transparency and automation across multiple stakeholders in SAGIN is utilizing blockchain-based smart contracts and distributed ledger technologies.

			\section{Conclusion} 
			This paper provides an overview of SAGIN-enabled MC and the critical challenges that arise from unifying heterogeneous space, air, and ground links. It also discussed the potential of AI, particularly agentic RL in enabling intelligent, adaptive SAGIN-enabled MC. Through a case study, the paper proposes an agentic RL-based algorithm that jointly optimizes link selection with respect to capacity, latency, and power. Simulation results showed that the proposed algorithm effectively balances competing performance objectives and outperforms multiple baselines. Lastly, this article presented key open research problems, which offer a roadmap toward scalable, resilient, and high-performance SAGIN-enabled MC.

			\section*{Acknowledgment}
			This work was supported by the National Research Foundation of Korea (NRF) grant funded by the Korean government
			(MSIT) (No. 2019R1proposed2007037 and 2022R1A4A3033401) and by the MSIT (Ministry of Science and ICT), Korea, under the ITRC
			(Information Technology Research Center) support program (IITP-2021-0-02046) supervised by the IITP
			(Institute for Information $\&$ Communications Technology Planning $\&$ Evaluation).
			This work was also supported in part by the SNS JU European Union’s Horizon Europe Research and Innovation Program under Grant 101139194 (6GXCEL); in part by the Horizon Europe Program under the MSCA Staff Exchanges under Grant 101086218 (EVOLVE); in part by FWO, Belgium, under Grant for a scientific stay in Flanders (file number V507025N).
			
			\balance

			\bibliographystyle{IEEEtran}
			\bibliography{IEEEabrv, ./LatexInclusion/SAGIN_MC}
			
			\begin{IEEEbiographynophoto}{Abd Ullah Khan}
				is a Researcher at ID Lab, Ghent University, Belgium. He is also affiliated with CQI Lab, Kyung Hee University, South Korea, and National University of Sciences and Technology, Pakistan.
			\end{IEEEbiographynophoto}
			
			\begin{IEEEbiographynophoto}{Adnan Shahid}
				is an Assistant Professor at imec-IDLab, Ghent University. His research interests include resource optimization and interference management in 5G/6G networks.
			\end{IEEEbiographynophoto}
			
			\begin{IEEEbiographynophoto}{Haejoon Jung}
				is a Professor at the Department of Electronic Engineering, Kyung Hee University.
			\end{IEEEbiographynophoto}
			
			\begin{IEEEbiographynophoto}{Hyundong Shin}
				is a Professor at Kyung Hee University, Korea. He received the IEEE Communications Society's Guglielmo Marconi Prize Paper Award and William R. Bennett Prize Paper Award. He was an Editor of IEEE Transactions on Wireless Communications and IEEE Communications Letters.
			\end{IEEEbiographynophoto}
			
		\end{document}